\pdfoutput=1 
\documentclass[prd,twocolumn,nofootinbib,preprintnumbers,superscriptaddress,showpacs]{revtex4-1}
\usepackage{xcolor}
\usepackage{graphicx}
\usepackage{bm}
\usepackage{amsmath}
\usepackage{lineno}
\usepackage{amssymb}
\usepackage[utf8]{inputenc}
\usepackage{multirow}
\usepackage{booktabs}
\usepackage[T1]{fontenc}
\usepackage{lipsum}
\usepackage{stmaryrd}
\usepackage{xcolor}
\usepackage{caption,subcaption}
\usepackage{multirow}
\usepackage{amssymb}
\usepackage{multirow}
\usepackage{booktabs}
\usepackage{siunitx}
\usepackage{tikz}
\usetikzlibrary{cd}
\tikzcdset{
arrow style=tikz,
diagrams={>={Straight Barb[scale=0.8]}}
}

\usepackage[colorlinks,pdfstartview=FitH]{hyperref}
\hypersetup{linkcolor=blue,citecolor=blue,filecolor=black,urlcolor=blue}

\IfFileExists{dsfont.sty}
	{\usepackage{dsfont}
         \let\mathbb=\mathds
         \newcommand{\id}{\mathds{1}}}
	{\typeout{Package dsfont.sty was not found, using alternative macros.}
         \let\mathds=\mathbb
         \newcommand{\id}{\mbox{1 \kern-.59em {\rm l}}}}

\DeclareMathOperator*{\argmax}{argmax}
\DeclareMathOperator*{\argmin}{argmin}

\newcommand\lb{\left(}
\newcommand\rb{\right)}

\renewcommand{\part}{{\rm part}}

\newcommand{\be}{\begin{equation}}
\newcommand{\ee}{\end{equation}}
\newcommand{\bes}{\begin{subequations}}
\newcommand{\ees}{\end{subequations}}
\newcommand{\bea}{\begin{eqnarray}}
\newcommand{\eea}{\end{eqnarray}}

\newcommand{\simpz}{ \left \langle \boldsymbol{\theta}_i  \right \rangle }

\newcommand{\simpo}{\left \langle \boldsymbol{\theta}_i  \boldsymbol{\theta}_j \right \rangle }

\newcommand{\bs}{\boldsymbol}

\def\nbox#1#2{\vcenter{\hrule \hbox{\vrule height#2in
\kern#1in \vrule} \hrule}}
\def\sq{\,\raise.5pt\hbox{$\nbox{.10}{.10}$}\,}
\def\sqb{\,\raise.5pt\hbox{$\overline{\nbox{.09}{.09}}$}\,}

\begin{document}
\author{Tejes Gaertner}
\email{tejes.gaertner@st-hughs.ox.ac.uk}
\affiliation{Mathematical Institute, University of Oxford, Oxford, OX2 6GG, U.K.}
\affiliation{Department of Physics and Astronomy, University of California, Los Angeles,CA 90095, U.S.A.}
\author{Jared Reiten}
\email{jdreiten@physics.ucla.edu}
\affiliation{Department of Physics and Astronomy, University of California, Los Angeles,CA 90095, U.S.A.}
\affiliation{Mani L. Bhaumik Institute for Theoretical Physics,  University of California,  Los Angeles,  CA 90095, U.S.A.}

\title{The simplicial substructure of jets}

\begin{abstract}
In this work, we construct a new data type for hadronic jets in which the traditional point-cloud representation is transformed into a simplicial complex consisting of vertices, or 0-simplexes. An angular resolution scale, $r$, is then drawn about each vertex, forming balls about hadrons. As $r$ grows, the overlap of balls form 2- and 3-point connections, thereby appending 1- and 2-simplexes to the complex. We thus associate a jet with an angular-resolution-dependent characterization of its substructure---we dub this data type the simplicial substructure complex $K_{\rm sub}(r)$. This data type gives rise to two interesting representations. First, the subset of 0-and 1-simplexes lends itself naturally to a graph representation of a given jet's substructure and we provide examples of valuable graph-theoretic calculations such a representation affords. Second, the subset of 1- and 2-simplexes gives rise to what is known as a Face-Counting-Vector, in topological combinatorics parlance. We explore information-theoretic aspects of the components of this vector, various metric properties which follow, as well as how this vector can be used to define new jet-shape observables. The utility of these representations is demonstrated in the context of the discrimination of jets initiated by light quarks and gluons from those initiated by tops.
\end{abstract}

\maketitle

\section{Introduction}
The field of jet substructure concerns itself primarily with understanding the complexity of the radiation patterns developed by partons emerging from high-energy particle collisions \cite{Larkoski:2017jix}. As such, it has proven itself to be a fertile ground for the application of various statistical and Machine-Learning-(ML)-based tools. For a comprehensive catalogue of such applications, see \cite{Feickert:2021ajf}. Early on, it was shown that jets and their substructure can be intuitively visualized as images \cite{Cogan:2014oua, deOliveira:2015xxd}, just as a calorimeter ``views'' them. Using this data type, one must perform a sequence of pre-processing steps to get the image in a standard format so that machines can learn the subtle differences among classes of jets, typically categorized according to the flavor of their initiating parton \cite{Komiske:2016rsd, Kasieczka:2017nvn}. Such pre-processing amounts to the application of isometries in the $(\eta,\phi)$ plane, physically corresponding to Lorentz boosts along the beam pipe. See \cite{Romero:2023hrk} for a recent work that encodes the rotational symmetry exhibited by jets into neural network architectures.

Another important direction is the application of concepts from Optimal Transport \cite{Villani:2009} to define a natural metric on the space of jets---see \cite{Komiske:2019fks, Komiske:2020qhg} for the original ideas and \cite{Cai:2020vzx, Cai:2021hnn} for extensions. Such a metric can be defined for jets in the discretized image format or the infinitesimal form known as a point-cloud. However, both representations still depend on the way in which jets are pre-processed, causing the ``distance'' between jets to inherit such dependence. Such an inheritance is ameliorated through use of an alternative data type, that is the spectral function \cite{Larkoski:2023qnv}, closely related to energy-energy correlators \cite{Basham:1978bw, Tkachov:1995kk, Jankowiak:2011qa, Larkoski:2013eya}. This data type makes use of inter-particle distances, thereby making its affiliated metric invariant under any pre-processing scheme. 

Yet another interesting avenue is the application of concepts from Topological Data Analysis \cite{chazal2021introduction} to the study of jet substructure. This is carried out in \cite{Li:2020jdb, yale-group}, where jets are analyzed through topological notions of connectivity, such as persistent homology---see \cite{Lim:2020igi, Hamilton:2022blu, Beuria:2023hkm} for other applications of such notions to the physics of high-energy particle collisions. The basic structure underlying such analysis is that of a simplicial complex---that is a collection of points, lines, and triangles, or equivalently 0-, 1-, and 2-simplexes, where points represent a particle with no connections, lines for two-particle connections, and triangles for three. Connections are activated based on the overlap of closed balls drawn about particles in the $(\eta,\phi)$ plane, and thus depend solely on inter-particle distances, making a simplical complex manifestly pre-processing independent. What is particularly novel about this construction is that a simplicial complex furnishes a one-parameter family of representations of a given jet, one for each value of angular resolution scale $r$---the radius of the ball drawn about each jet constituent.

The aim of this work is to unify the two aforementioned approaches to jet substructure---that is to represent a jet in a data type that (1) can be naturally endowed with a metric, (2) is completely independent of any choice of pre-processing, and (3) is a one-parameter family of representations, thus forming a thorough tomographic view of the various scales which characterize its substructure. We do this through defining a new fundamental data type, which we refer to as the simplical substructure complex of a jet, denoted $K_{\rm sub}(r)$. This construction leads to two interesting limiting representations for jets, the first being a graph \cite{Diestel:2017}, which we label as $G_{\rm sub}(r)$ and enables the computation of a host of graph-theoretical objects. The second representation is what is known as a Face-Counting-Vector ($f$-vector) \cite{Ziegler:1995}, denoted $\boldsymbol{f}(r)$, whose components are the number of 1- and 2-simplexes at a particular scale resolution scale $r$. The $f$-vector representation is particularly rich and the bulk of this work will be in the exploration of its features.

This paper is organized as follows. Sec.~\ref{sec:complex} will be devoted to developing the data type and its limits, with Sec.~\ref{sec:simplex} dealing with the construction of the simplicial substructure complex $K_{\rm sub}(r)$, Sec.~\ref{sec:graph} the limiting case of the substructure graph $G_{\rm sub}(r)$, and Sec.~\ref{sec:f-vector} the $f$-vector $\boldsymbol{f}(r)$. In Sec.~\ref{sec:further-investigations} we examine geometric and information-theoretic features of the components of the $f$-vector---those dealing with their stochastic nature are collected in  Sec.~\ref{sec:variable-info}, while those in reference to their flavor-dependence (QCD vs. top jets) in Sec.~\ref{sec:flavor-info}. In Sec.~\ref{sec:jet-shape} we demonstrate how the $f$-vector leads naturally to the definition of new jet shape observables. We conclude in Sec.~\ref{sec:conclusion}.

\section{Jet substructure with simplicial complexes}
\label{sec:complex} 

In this section, we define the simplicial substructure complex for a jet, which we denote as $K_{\rm sub}(r)$, where $r$ is an angular scale that resolves the internal substructure of the jet. This is a fundamental object that naturally gives rise to two alternative jet representations, namely (1) a substructure graph, denoted $G_{\rm sub}(r)$ and (2) a substructure face-counting vector, $\boldsymbol{f}(r)$. These two representations may be viewed as particular limits of $K_{\rm sub}(r)$.

\subsection{The simplicial complexes for a jet: basic definitions}
\label{sec:simplex}

\begin{figure}
\begin{centering}
\includegraphics[width=0.8\linewidth]{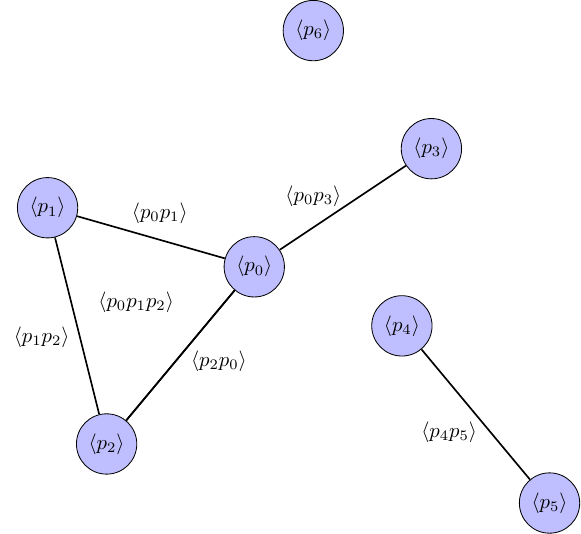}
\end{centering}
 \caption{An example of a simplicial complex, $\mathcal{C}$. Here we have that $S_{-1} = \emptyset$, $S_0 = \{ \langle p_0 \rangle, \dots, \langle p_6 \rangle \}$, $S_1 = \{ \langle p_0p_1 \rangle, \langle p_1p_2 \rangle, \langle p_2p_0 \rangle, \langle p_0p_3 \rangle, \langle p_4p_5 \rangle \}$, $S_2 = \{\langle p_0p_1p_2 \rangle \}$, and therefore $\mathcal{C} = \bigcup _{\ell = 0}S _{\ell - 1}$.}
 \label{fig:basic-simplex}
\end{figure}

The basic and natural data type for a jet is that of a point-cloud. A jet is a collection of the four-momentum vectors for the hadrons contained therein. The on-shell condition for final state particles then reduces the number of degrees of freedom from four down to three, which for hadron colliders like the LHC, are conveniently chosen to be the transverse momentum with respect to the beam pipe, $p_T$, together with the rapidity, $\eta$, and the azimuthal angle, $\phi$. Then, normalizing the transverse momentum of each particle by that of the overall value for the jet, we obtain the fractions $z_i \equiv p_{Ti}/p_T^{\rm jet}$. The three degrees of freedom for each hadron are then $(z_i, \eta_i, \phi_i)$. With these three degrees of freedom, we construct the point-cloud representation for a jet
\begin{align}
    \mathcal{J}_{\rm pc}(\eta,\phi) 
     = \sum_{i \in \mathrm{jet}}z_i \, \delta^{(2)} (\boldsymbol{\theta} - \boldsymbol{\theta}_i ) \,,
\end{align}
where $\boldsymbol{\theta}_i \equiv (\eta_i, \phi_i)$. Hence, the point-cloud format represents the jet as a discrete probability distribution in rapidity-azimuth space with $p_T$-fractions as weights.

In this work, the base from which we start will be a simpler object, namely, the collection of $\boldsymbol{\theta}$-vectors. We will refer to this collection as $\mathcal{J}_{\theta}$, and let us also have $\mathcal{I} = \{1,2,\dots,N_{\rm jet}\}$ denote the indexing set so that
\begin{align}
\label{eq:J-theta-rep}
    \mathcal{J}_{\theta} = \left\{\boldsymbol{\theta}_i \in \mathbb{R}^2 \right\}_{i \in \mathcal{I}} \,,
\end{align}
where $N_{\rm jet}$ is the multiplicity of the jet. Note we are thus considering the jet as subset of the effective $\mathbb{R}^2$ formed by the calorimeter cells onto which the final state hadrons deposit themselves. Further note, that in this construction, we have completely dropped any $z_i$-dependence, treating particles purely in terms of their positions. 

Next, for each particle in the jet, we can draw a closed ball about it, i.e. for the $i^{\rm th}$ particle, we construct the ball
\begin{align}
    B_i(r) \equiv \big\{ \boldsymbol{\theta} \in \mathbb{R}^2 \, | \, \lVert \boldsymbol{\theta} - \boldsymbol{\theta}_i \rVert \leq r \big\}\,,
\end{align}
where $\lVert \cdot \rVert$ denotes the Euclidean norm.

In what follows, since we are working with vectors in $\mathbb{R}^2$, we will consider simplexes up to order two, that is, 0-, 1-, and 2-simplexes. A 0-simplex will just be taken as a point $\boldsymbol{\theta}_i \in \mathbb{R}^2$ and denoted as $\left \langle \boldsymbol{\theta}_i \right \rangle$. A 1-simplex will be visualized as the line segment that connects two points $\boldsymbol{\theta}_i$ and $\boldsymbol{\theta}_j$ and denoted as $\left \langle \boldsymbol{\theta}_i \boldsymbol{\theta}_j\right \rangle$. Lastly, a 2-simplex can be visualized as the triangle formed by the connections between three points $\boldsymbol{\theta}_i$, $\boldsymbol{\theta}_j$, and $\boldsymbol{\theta}_k$, which is then denoted $\left \langle \boldsymbol{\theta}_i \boldsymbol{\theta}_j \boldsymbol{\theta}_k \right \rangle$.

The generation of simplexes will come about through the varying of the radius $r$ of the balls about each point in each jet uniformly. 1- and 2-simplexes are then spawned at the onset of nonempty intersections between pairs and triplets of balls, respectively. The number of 0-simplexes stays the same for all values of $r$, as it is just the set of vertices
\begin{align}
    S_0(r) = \left\{ \langle \boldsymbol{\theta}_i \rangle  \right\}_{i \in \mathcal{I}}\,.
\end{align}
As $r$ grows, connections proliferate and one starts amassing collections of 1- and 2-simplexes. The set of the former is denoted
\begin{align}
    S_1(r) = \left\{ \langle \boldsymbol{\theta}_i \boldsymbol{\theta}_j \rangle \, \big| \, B_i(r) \cap B_j(r) \neq \emptyset \right\}_{i,j \in \mathcal{I}}\,,
\end{align}
where $\emptyset$ is the empty set. Thus, the condition for a connection between two points can be equivalently stated as
\begin{align}
\label{eq:argmin-1-2}
    B_i(r) \cap B_j(r) \neq \emptyset  \iff r \geq \frac{\lVert \boldsymbol{\theta}_i - \boldsymbol{\theta}_j \rVert}{2}\,.
\end{align}
The set of 2-simplexes is then
\begin{align}
    S_2(r) = \left\{ \langle \boldsymbol{\theta}_i \boldsymbol{\theta}_j \boldsymbol{\theta}_k \rangle \, \big| \, B_i(r) \cap B_j(r) \cap B_k(r) \neq \emptyset \right\}_{i,j,k \in \mathcal{I}}\,.
\end{align}
For purposes of formality, let us define one last trivial set 
\begin{align}
    S_{-1} = \emptyset \,.
\end{align}
Equipped with these sets, the simplicial substructure complex for a given jet is then defined as
\begin{align}
\label{eq:K-sub}
    K_{\rm sub}(r) \equiv \bigcup _{\ell = 0}^3 S_{\ell-1} (r) \,.
\end{align}

At this point, an important physical comment is in order. Note that the simplicial substructure complex of Eq.~(\ref{eq:K-sub}) is purely a function of the radial coordinate $r$, whose magnitude is determined by the relative angular separations in the calorimeter cell. As such, $K_{\rm sub}(r)$ is naturally invariant under isometries of the plane. $\mathrm{ISO}\left( \mathbb{R}^2\right)$-invariance is to say invariance under translations, reflections, and rotations in $(\eta,\phi)$-space. From a physical standpoint, such transformations correspond to Lorentz boosts along the beam pipe (canonically taken to define the $z$-axis) and from a data-formatting viewpoint, this is to say that $K_{\rm sub}(r)$ is independent of pre-processing choice \cite{Cogan:2014oua, deOliveira:2015xxd}. 

Before proceeding, we describe the particular data set used in this work, as well as the jet-trimming procedure we preform on the jets contained therein. The data set is the standard one used for the benchmarking of top-tagging architectures \cite{Kasieczka:2019dbj}. The set consists of signal top jets amongst a mixed background of jets initiated by light quarks and gluons. These jets are simulated with Pythia8 \cite{Bierlich:2022pfr} for collisions at $\sqrt{s}= 14$ TeV. Detector effects are simulated with Delphes \cite{deFavereau:2013fsa} with the ATLAS card. Jets of radius $R=0.8$ are then clustered using the anti-$k_T$ algorithm \cite{Cacciari:2008gp} through FastJet \cite{Cacciari:2011ma}. Each jet in the data set is the leading jet of the event from whence it came and falls in the transverse momentum range of $p_T^{\mathrm{jet}} \in [550, 650]$ GeV. Jets in this data set contain around fifty or so particles each---the vast majority of these being exceedingly low in transverse momentum and hence capturing soft physics. In the entirety of our proceeding analysis, we implement a basic trimming procedure \cite{Krohn:2009th} in order to capture the hard/collinear physical scales which dominate the aspects of substructure in which QCD and top jets differ. Our procedure consists of simultaneously implementing a particle transverse momentum cut of $p_T^{\mathrm{cut}} \approx 3$-$5$ GeV as well as a multiplicity cut of $N=10$. Thus, we order the particles in descending order of their transverse momentum and keep only the ten highest-$p_T$ particles, so long as they are above $p_T^{\mathrm{cut}}$. This guarantees that our jets are not contaminated by soft physics as well as gives our jets a fixed value of $N_{\mathrm{jet}} = 10$, which, as we will see, allows us to make sharp statements regarding various limits. Furthermore, we find this multiplicity cut to be quite ideal in revealing the critical angular resolution scale over which QCD and top jets differ the most, that is, what we refer to as the characteristic decay angle of tops
\begin{align}
    \theta_{\mathrm{cd}} \equiv \frac{m_{\mathrm{top}}}{p_T^{\mathrm{jet}}}\,,
\end{align}
where $m_{\mathrm{top}}$ is the top mass.\footnote{Note that this is technically the same as the ``dead-cone'' angle $\theta_{\rm dc}$, however that nomenclature should be reserved for the suppression of soft gluon radiation off a heavy quark. This is notoriously difficult to measure for top quarks \cite{Maltoni:2016ays} due to their short lifetime. However, it is identical to the characteristic opening angle of the decay products of boosted heavy particles.} This is the unique angular resolution scale that can be constructed purely out of the available mass-dimension-1 parameters of the data set. We will see that the $r$-dependence of $K_{\mathrm{sub}}(r)$ is inherently what allows $\theta_{\mathrm{cd}}$ to reveal itself in such a wide variety of contexts as the critical scale of our data set.

\subsection{The graph representation of a jet}
\label{sec:graph}

We can think of the simplicial substructure complex, given by Eq.~(\ref{eq:K-sub}), as a principal representation of a jet, through which there exist limits which give rise to unique ways to resolve the substructure of a jet. In this section, we will consider the particular limit that restricts us to the set of 0- and 1-simplexes. Doing so will give rise to an $r$-dependent graph representation for each jet \cite{Diestel:2017}, which we will refer to as $G_{\rm sub}(r)$. We begin with some graph-theoretic preliminaries.

A weighted graph is a collection of vertices, edges, and edge weights $G=(V, E, W)$ where $V$ is the set of vertices, $E$ the set of edges, and $W$ the set of edge weights. Using the language developed in the previous section, we can understand a weighted graph as a particular limit of a simplicial complex. In particular, we can identify each vertex with a 0-simplex and each edge with a 1-simplex, thus
\begin{align}
    V(r) \equiv S_0(r) \,, \\
    E(r) \equiv S_1(r) \,,
\end{align}
where we make explicit the fact that both $V$ and $E$ inherit $r$-dependence from our simplicial construction. Next, let us consider a weight function, $\omega$, that applies a weight to each edge according to
\begin{align}
\omega \colon  E(r) &\longrightarrow \mathbb{R}^{\geq 0}\,, \nonumber \\
\simpo &\longmapsto  \left \lVert \boldsymbol{\theta}_i - \boldsymbol{\theta}_j \right \rVert \,.
\end{align}
The weight set $W$ can thus be understood as the image of the edge set $E$ under $\omega$, and hence inherits $r$-dependence therefrom: $W(r) = \omega(E(r))$.

With these three sets so defined, we see that the simplicial substructure complex naturally gives rise to the graph substructure complex
\begin{align}
    K_{\rm sub}(r) \longrightarrow G_{\rm sub}(r)\,.
\end{align}
At this point we reiterate that as $r$ grows, the development of 0- and 1-simplexes gives rise to non-trivial connection structures within each individual jet. A key distinction between the $G_{\rm sub}(r)$ representation and the reduced $\mathcal{J}_\theta$ representation is that for each pair $\bs{\theta}_i,\bs{\theta}_j \in \mathcal{J}_\theta$, there is inherently a connection between them defined by their Euclidean distance $\lVert \bs{\theta}_i - \bs{\theta}_j \rVert$, whereas for the corresponding pair $\langle \bs\theta_i \rangle, \langle \bs\theta_j \rangle \in G_{\rm sub}(r)$, the same connection of $\lVert \bs{\theta}_i - \bs{\theta}_j \rVert$ between them is only ascribed as a weight once the edge $\simpo$ is formed, which is to say only once $r \geq\lVert \bs{\theta}_i - \bs{\theta}_j \rVert/2 $. This leads to a non-trivial evolution of connection patterns in the graph representation. Such connections can be characterized by defining a path metric on the graph, where a path is taken to be a sequence of 0-simplexes that are connected through 1-simplexes:
\begin{align}
\label{eq:path-metric}
    &d_{\rm path}\left(\left \langle \boldsymbol{\theta}_i  \right \rangle,\left \langle \boldsymbol{\theta}_j  \right \rangle \right) = \min \biggl\{ \sum_{k=0}^{n-1}\omega\lb \left \langle \boldsymbol{\theta}_{A_k} \boldsymbol{\theta}_{A_{k+1}} \right \rangle   \rb \, \bigg| \,   \nonumber \\
    &\quad \left \langle \boldsymbol{\theta}_{A_k} \boldsymbol{\theta}_{A_{k+1}} \right \rangle \in S_1(r) \,,\, \left \langle \boldsymbol{\theta}_{A_0} \right \rangle = \left \langle \boldsymbol{\theta}_i \right \rangle\,,\, \left \langle \boldsymbol{\theta}_{A_n} \right \rangle = \left \langle \boldsymbol{\theta}_j \right \rangle \biggr\}\,.
\end{align}

\begin{figure*}[htb]
    \centering 
\begin{subfigure}{0.4\textwidth}
  \includegraphics[width=\linewidth]{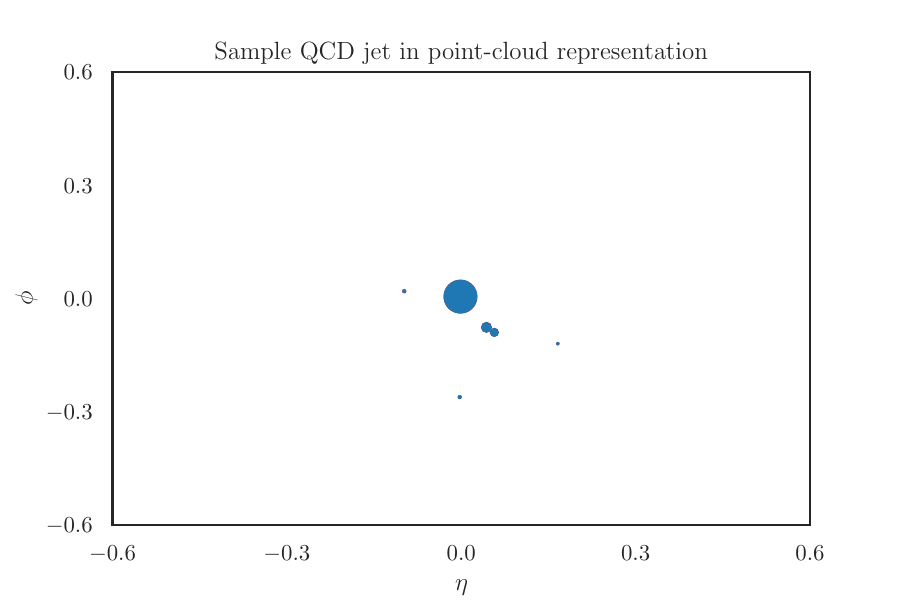}
  \label{fig:qcd-cloud}
\end{subfigure}\hfil 
\begin{subfigure}{0.4\textwidth}
  \includegraphics[width=\linewidth]{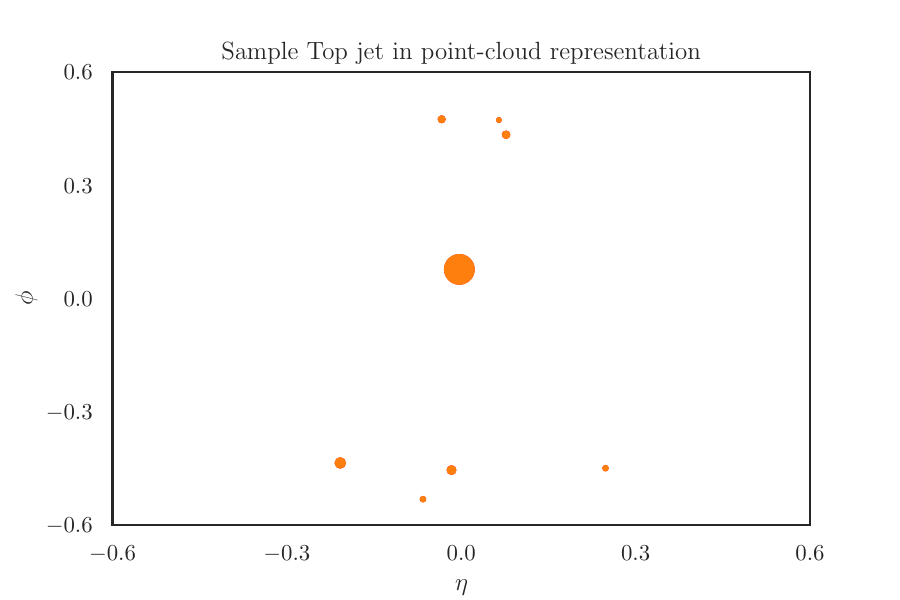}
  \label{fig:top-cloud}
\end{subfigure}\hfil 

\medskip
\begin{subfigure}{0.4\textwidth}
  \includegraphics[width=\linewidth]{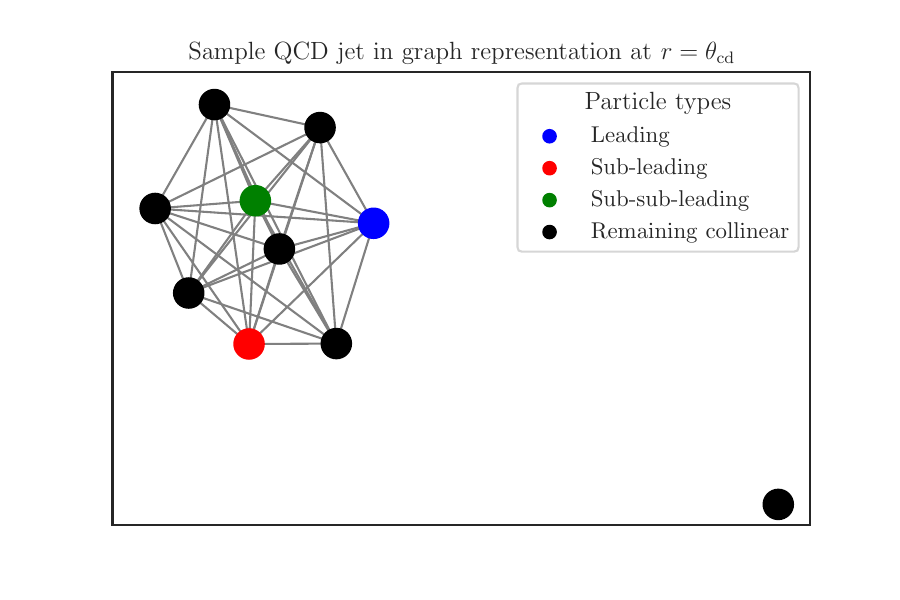}
  \label{fig:qcd-graph}
\end{subfigure}\hfil 
\begin{subfigure}{0.4\textwidth}
  \includegraphics[width=\linewidth]{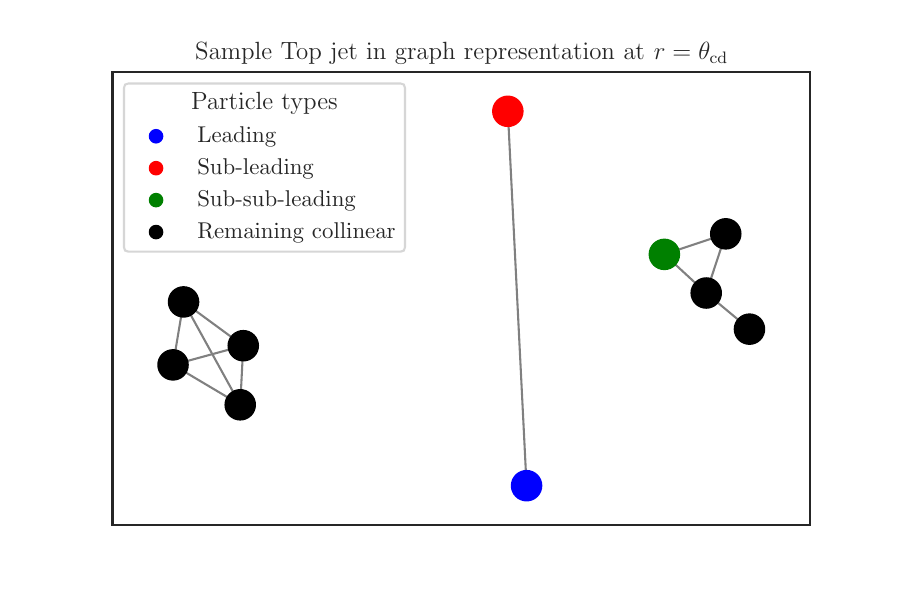}
  \label{fig:top-graph}
\end{subfigure}\hfil 
\caption{Examples of QCD (left) and top (right) jets in the point-cloud (top panels) and graph (bottom panels) representations. Note that in each case, our trimming procedure has already been enacted, leaving both jets with ten particles each. The graph representation is taken at the characteristic angular scale $r \simeq m_{\mathrm{top}}/p_T^{\mathrm{jet}}$. At this level of resolution, QCD and top graphs differ substantially in their both their numbers of connected components as well as the number of nodes contained therein.}
\label{fig:qcd-top-representations}
\end{figure*}

With this metric in hand, we can ask questions regarding the global connectivity structure of a jet, and how this structure evolves with $r$. One particular quantity that is defined on the set of vertices $V(r)$ is known as the closeness centrality
\begin{align}
    C_B(\left \langle \boldsymbol{ \theta} \right \rangle) = \left(\frac{1}{N_{\rm jet}}\sum_{\left \langle \boldsymbol{\theta}_i \right \rangle \sim \left \langle \boldsymbol{\theta} \right \rangle} d_{\rm path}\left(\left \langle \boldsymbol{\theta}  \right \rangle,\left \langle \boldsymbol{\theta}_i  \right \rangle \right) \right)^{-1}\,,
\end{align}
which can be interpreted as the inverse of the average distance between the 0-simplex $\langle \bs\theta \rangle$ and all other 0-simplexes that it is path-connected to. Therefore, a smaller average distance indicates greater proximity of $\langle \bs\theta \rangle$ to all other points, positioning it more centrally within the entire graph.

In order to affiliate a single value of $C_B$ with a jet, we consider the mean value taken over all vertices
\begin{align}
    \overline{C_B}(r) = \sum_{\simpz \in G_{\mathrm{sub}}(r)}C_B(\simpz)\,.
\end{align}

\begin{widetext}

\begin{figure}
\begin{centering}
\includegraphics[width=0.45\linewidth]{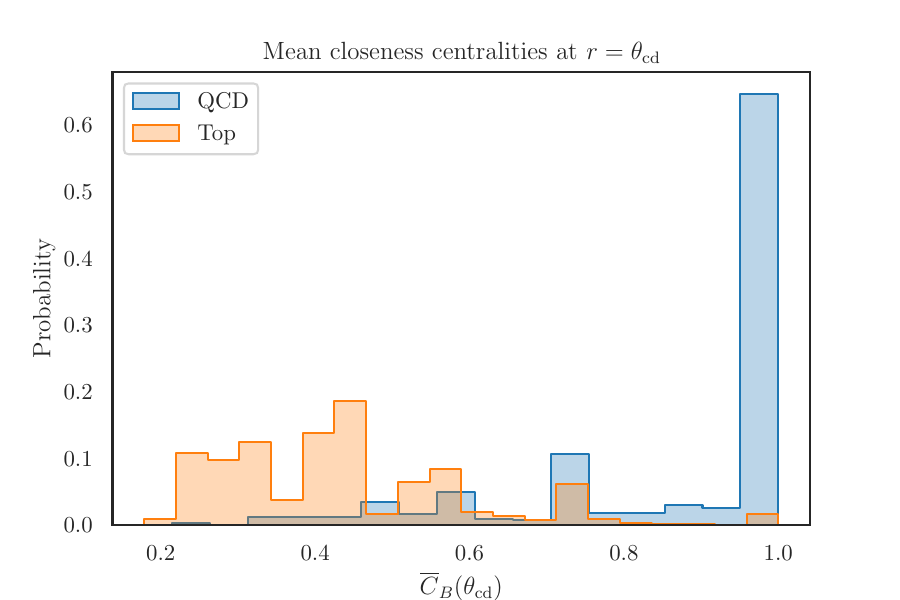}
\includegraphics[width=0.45\linewidth]{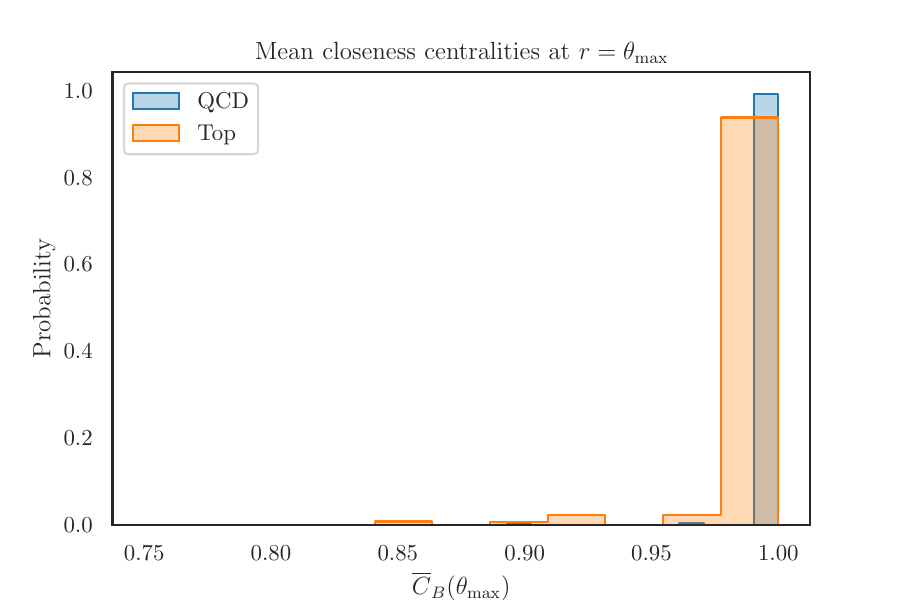}
\end{centering}
\caption{Closeness centrality distributions for QCD (blue) and top (orange) jet graphs evaluated at two angular resolution scales. The left plot is evaluated at the characteristic decay angle $r = \theta_{\mathrm{cd}}$ while the right is evaluated at the maximal angular resolution scale $r = \theta_{\mathrm{max}}$. We see that the behavior of closeness centrality distributions is largely dependent on the connectivity of the underlying graph, as at $r = \theta_{\mathrm{max}}$ both QCD and top jets are fully-connected and have their distributions collapse nearly onto one-another. This is to be contrasted with the case at $r = \theta_{\mathrm{cd}}$, where clear separability is achieved between the distributions.}
 \label{fig:centralities}
\end{figure}

\end{widetext}

Distributions for these mean closeness centralities for QCD and top jets for two different levels of angular resolution are depicted in Fig.~\ref{fig:centralities}. In this figure, we see that at the characteristic scale for the tops, QCD jets by and large exhibit higher values than top jets, which is to say that the center of a QCD jet is predominantly close to its neighbors, whereas that of a top jet is, on average, much further from its neighbors. This is a graph-theoretic way of probing the following physical fact: the top jets have characteristic decay products which are boosted and spread out, on top of which QCD evolution takes place, adorning each prong with a haze of radiation. Thus, the center of the jet is inevitably far from all the haze spawned by each of the different decay products. Since QCD jets are predominantly single-pronged, the resulting haze surrounds this prong and the center itself. Thus, the mean closeness centrality evaluated at $r=\theta_{\mathrm{cd}}$ is able to capture the connected/disconnected structure depicted in Fig.~\ref{fig:qcd-top-representations} for QCD/top graphs, respectively. This is confirmed by the distributions resulting from an angular resolution of $r=\theta_{\mathrm{max}}$, where both QCD and top jets achieve full inter-connectivity among their vertices, and collapse to the same high value of $\overline{C_B}$.

To complement the global connectivity structure related to the elements of $S_0(r)$, we next look to the local structure that we may infer from elements of $S_1(r)$. To do so, we compute the Ricci curvature for a graph, known as the Forman curvature \cite{Sreejith_2016}, which is defined on the set of 1-simplexes as 
\begin{align}
\label{eq:forman-def}
    &R_{F}\left(\left \langle \boldsymbol{\theta}_i\boldsymbol{\theta}_j \right \rangle \right)   \nonumber \\
    &\quad = 2 -\sum_{k \sim i,\ell \sim j} \left(\sqrt{ \frac{\omega\left(\left \langle \boldsymbol{\theta}_i\boldsymbol{\theta}_j \right \rangle \right)}{\omega\left(\left \langle \boldsymbol{\theta}_i\boldsymbol{\theta}_k \right \rangle \right)} } + \sqrt{\frac{\omega\left(\left \langle \boldsymbol{\theta}_i\boldsymbol{\theta}_j \right \rangle \right)}{\omega\left(\left \langle \boldsymbol{\theta}_j\boldsymbol{\theta}_\ell \right \rangle \right) }} \right)\,,
\end{align} 
where the sum $k\sim i$ and $\ell \sim j$ are over immediate neighbors $k$ of $i$ and $\ell$ of $j$. This is to say paths that include only a single segment, which reduces $d_{\rm path}\left(\left \langle \boldsymbol{\theta}_i  \right \rangle,\left \langle \boldsymbol{\theta}_j  \right \rangle \right) \rightarrow \omega(\langle \bs\theta_i \bs\theta_j \rangle)$, give this particular quantity a notion of locality.

Just as we did with the closeness centrality, we ascribe to each jet its mean value of Forman curvature, taken over all edges in each $G_{\mathrm{sub}}(r)$:
\begin{align}
    \overline{R_F} = \sum_{\simpo \in G_{\mathrm{sub}}(r)} R_F(\simpo)\,.
\end{align}

\begin{widetext}

\begin{figure}
\begin{centering}
\includegraphics[width=0.45\linewidth]{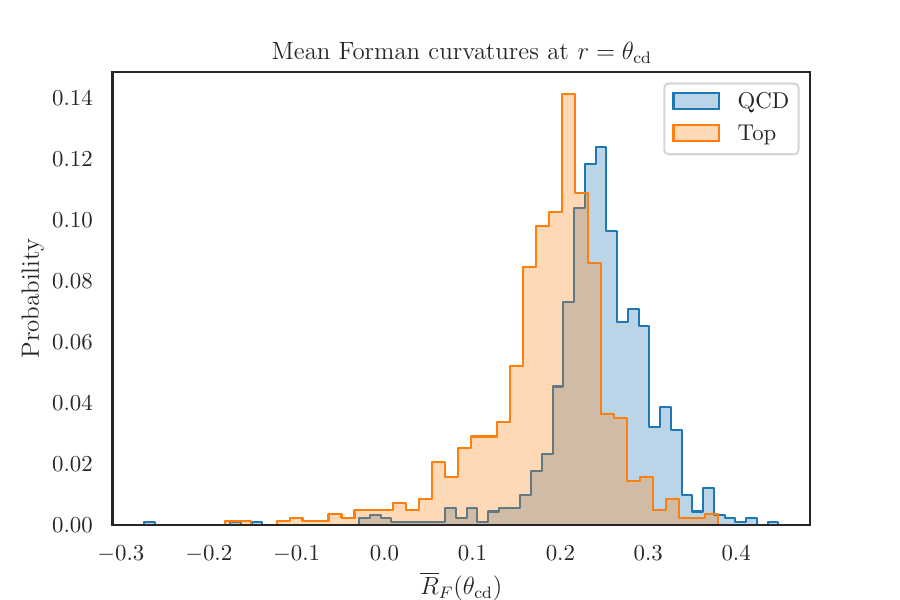}
\includegraphics[width=0.45\linewidth]{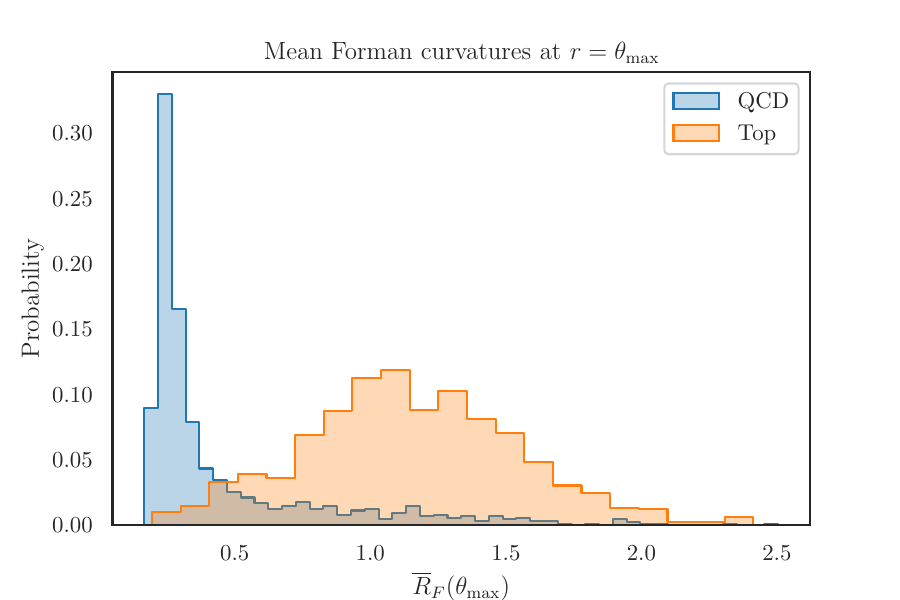}
\end{centering}
 \caption{Forman curvature distributions for QCD (blue) and top (orange) jet graphs evaluated at two angular resolution scales. The left plot is evaluated at the characteristic decay angle $r = \theta_{\mathrm{cd}}$ while the right is evaluated at the maximal angular resolution scale $r = \theta_{\mathrm{max}}$. We see that the behavior of the Forman curvature distributions is somewhat opposite to those of the closeness centrality in that QCD and top distributions nearly coincide at scale $r = \theta_{\mathrm{cd}}$ while they are highly separable at $r = \theta_{\mathrm{max}}$. This is a manifestation of the Forman curvature local nature, which only differentiates top from QCD jets when the entirety of the extended structure of its decay products is resolved.}
 \label{fig:formans}
\end{figure}

\end{widetext}

Fig.~\ref{fig:formans} displays distributions of the mean Forman curvatures for QCD and top jets at the same two angular resolution scales as in Fig.~\ref{fig:centralities}---namely $r = \theta_{\mathrm{cd}}$ and $r = \theta_{\mathrm{max}}$. From this plot, we see that the mean Forman curvature behaves in a way that is inverse with respect to angular resolution scale to the behavior of the mean closeness centrality. We can understand this to arise from the sum appearing in Eq.~(\ref{eq:forman-def}), which is taken over only the immediate neighbors of vertices. Again, through the intuition garnered in Fig.~\ref{fig:qcd-top-representations}, at $r = \theta_{\mathrm{cd}}$ the disconnected structure of the top jets essentially collects three prongs over which to obtain mean curvatures from, each prong capturing a beam of QCD radiation, and therefore cause the top distribution to take on values near those of the QCD. Conversely, once we reach $r=\theta_{\mathrm{max}}$, all vertices gain connections, and therefore all vertices become immediate neighbors to all others. Thus, the sum over neighbors appearing in Eq.~(\ref{eq:forman-def}) becomes saturated. In going from $\theta_{\mathrm{cd}}$ to $\theta_{\mathrm{max}}$, few neighbors are added to this summation in the case of QCD jets, thus resulting in only a minor additional spread to its distribution. This is to be contrasted with the case of top jets, whose vertices not only acquire far more neighbors over which to sum, but also the large edge weights which come with these very neighbors, characteristic of the far-extended angular substructure of top jets relative to QCD.

We thus see that the jet representation furnished by $G_{\mathrm{sub}}(r)$ affords us the ability to compute graph-theoretic quantities, such as the closeness centrality and Forman curvature, which in turn, shed light on local/global features of the connectivity structure of jets probed at differing levels of angular resolution. We hope this graph representation opens the door to the exploration of many other graph-theoretic concepts which can provide a helpful perspective from which to view jet substructure.

\subsection{The $f$-vector representation of a jet}
\label{sec:f-vector}

As alluded to previously, various limits of the simplicial substructure complex $K_{\rm sub}(r)$ highlight different aspects probed by the angular resolution scale $r$. The last section dealt with aspects of 0- and 1-simplexes, while the purpose of this section is to shed light on important features regarding the 1- and 2-simplexes. An obvious feature common to both $S_1(r)$ and $S_2(r)$ is that their cardinalities are monotonically-increasing functions of $r$, whereas that of $S_0(r)$ is constant and thus independent of $r$. To characterize such aspects, we define the cardinalities of each subset of simplexes forming $K_{\rm sub}(r)$:
\begin{align}
\label{eq:f-def}
    f_\ell (r) \equiv \big| S_\ell (r) \big| \,,
\end{align}
where, by convention, we take $f_{-1}(r) = 1$ and we see trivially that $f_0(r) = N_{\rm jet}$.

This definition gives rise to the construction of what is known as the face-counting vector \cite{Ziegler:1995}, or $f$-vector for short, for a given complex:
\begin{align}
\label{eq:f-vec4}
    \boldsymbol{f}(r) = \big(f_{-1}(r), f_0(r), f_1(r), f_2(r)\big) \in \mathbb{N}^4\,.
\end{align}
The components of the $f$-vector can equivalently be understood as arising from the following generating function, known as the $f$-polynomial\footnote{
There exists a related polynomial, referred to as the $h$-polynomial
\begin{align}
    h(t) = \sum_{\ell = 0}^d h_\ell \,t^\ell\,, \nonumber
\end{align}
where the two are related via
\begin{align}
    \sum _{\ell = 0}^d f_{\ell-1}\,(t-1)^{d-\ell} = \sum _{\ell = 0}^d h_\ell \, t^{d-\ell}\,. \nonumber
\end{align}
Upon expanding the above expression, one may identify $h_3$ with the reduced Euler Characteristic 
\begin{align}
    h_3 &= -f_{-1} + f_0 - f_1 + f_2\,, \nonumber \\
    &= -1 + V - E + F \,, \nonumber \\
    &= -1 + \chi\,, \nonumber \\
    &= \chi_{\rm red}\,, \nonumber
\end{align}
where $V$ is the number of vertices ($f_0$), $E$ is the number of edges ($f_1$), and $F$ is the number of faces ($f_2$) in the classical notation of polyhedra; $\chi$ being the usual Euler characteristic. 

One may also extract the reduced Euler characteristic from the Euler-Poincar\'e formula for the $f$-vector
\begin{align}
    \chi_{\rm red} &= \sum_{\ell = 0}^d(-1)^{\ell - 1} f_{\ell-1}\,. \nonumber
\end{align}
By the same token, the $h$-polynomial implies the existence of the $h$-vector of its coefficients.
}
\begin{align}
    f(t) = \sum_{\ell = 0}^d f_{\ell - 1}\, t^\ell\,.
\end{align}
Note that by studying jets of fixed multiplicity we automatically know the maximum cardinalities of each simplex set $S_\ell$, which is to say we know the limits as $r\gtrsim R$ where $R$ is the overall jet radius. The limit is given simply by
\begin{align}
    \lim _{r \gtrsim R} S_{\ell - 1}(r) &= \binom{N_{\rm jet}}{\ell} \,.
\end{align}

This brings us to the key feature of this particular limit of $K_{\mathrm{sub}}(r)$. By normalizing all the jets in a data set to have the same fixed multiplicity---according to the procedure defined previously---we see that the $f$-vector of Eq.~(\ref{eq:f-vec4}) has two constant components and can thus be dimensionally-reduced to
\begin{align}
    \bs f(r) \rightarrow \left(f_1(r), f_2(r) \right) \in \mathbb{N}^2 \hookrightarrow \mathbb{R}^2\,.
\end{align}
We will refer to this representation as the Simplicial Face-Counting Vector, or SFV for short. Note that by considering $\mathbb{N}^2$ to be embedded in $\mathbb{R}^2$, we can define the metric on the $f$-vectors to simply be that of the Euclidean metric induced by $\mathbb{R}^2$ under this embedding. As such, we can immediately visualize a sample of jets as living in the effective plane, which we denote as 
$\mathbb{R}^2_{\rm S}$. As $r$ evolves, $\bs f(r)$ changes the locations of jets in $\mathbb{R}^2_{\rm S}$, and such motion can be visualized in Fig.~\ref{fig:qcd-top-kde}. In describing this emergent space, we will use notation $\mathbb{R}^2_{\rm S} = \mathbb{F}_1\times \mathbb{F}_2$ so that $(f_1(r), f_2(r)) \in \mathbb{F}_1\times \mathbb{F}_2\equiv \mathbb{R}^2_S \subset \mathbb{R}^2$, in order to distinguish the respective subspaces occupied by each component. We remark that $\mathbb{R}^2_S \subset \mathbb{R}^2$ in the strict sense, as
\begin{align}
    \binom{n(r)}{2} < \binom{n(r)}{3}\,,
\end{align}
where $n(r)$ is the effective number of 0-simplexes giving rise to 1- and 2-simplexes at the scale $r$. This inequality implies that  $\mathbb{R}^2_S$ can be approximately intuited as the lower right triangular region contained in the rectangle spanned by 
\begin{align}
    f_1 \in \left[0, \binom{N_{\mathrm{jet}}}{2} \right] \,, \quad f_2 \in \left[0, \binom{N_{\mathrm{jet}}}{3} \right]\,.
\end{align}

Displayed in Fig.~\ref{fig:qcd-top-kde} are kernel-density-estimate (KDE) plots depicting the distributions of QCD and top jets in this new space at three different levels of angular resolution---each level of resolution made with reference to $\theta_{\mathrm{cd}}$. The lowest level is $\theta_{\mathrm{cd}}/4$, where the total number of 1- and 2-simplexes is far below their maximum value thus both QCD and top jets occupy the lower left region of the space. We note that the QCD jets are smeared out over a larger range of values due to their predominantly single-pronged nature, while conversely, the higher degree of localization as manifested by the tops is due to their multi-pronged structure. The highest of the three resolution scales is chosen to be $2\times \theta_{\mathrm{cd}}$, for which the vast majority of QCD jets begin to saturate their limits, while top jets are still smeared out over a larger expanse in their accumulation of simplexes. The intermediate scale is taken to be $\theta_{\mathrm{cd}}$ where we see, as is to be expected by now, the greatest separation between the two classes. At this characteristic angular scale of top jets, simplexes defined over their multiple prongs have largely saturated, and above this scale, simplexes defined across the various prongs begin to accumulate and the final approach to the limiting distribution can be made. This is to be contrasted with QCD jets, who by this point, have very-nearly saturated their limits. In Fig.~\ref{fig:f1-f2}, we replicate the projections onto the $\mathbb{F}_1$ and $\mathbb{F}_2$ spaces for a closer look at the separation between the distributions at the scale $\theta_{\mathrm{cd}}$. 

We remark that a noteworthy feature of this construction is the very fact that $\mathbb{R}^2_S \subset \mathbb{R}^2$ and thus our embedding is readily visualized. Due to this, we can certainly classify QCD and top jets accurately by simply employing a histogram cut by eye on either of the projections in Fig.~\ref{fig:f1-f2}. Though alternatively, our Euclidean embedding allows for us to employ the standard unsupervised clustering algorithm, known as K-means \cite{scikit-learn}, in order to tag tops from the QCD background in $\mathbb{R}^2_S$ at the scale $r=\theta_{\mathrm{cd}}$. By simply recognizing the fact that we have a clear bimodal distribution at this scale, we can initialize K-Means to search for two clusters. Denoting our pair of clusters as $\mathcal{C} = \{C_1, C_2\}$, this amounts to the minimization task
\begin{align}
    \argmin _{\mathcal{C}} \sum_{a=1}^2 \frac{1}{\left| C_a\right|} \sum_{\bs f_i,\bs f_j \in C_a} \left \lVert \bs f_i(\theta_{\rm cd}) - \bs f_j(\theta_{\rm cd}) \right\rVert^2\,,
\end{align}
which, upon completion, clusters jets with $84\%$ accuracy.\footnote{This accuracy is determined by the $F_1$ score, a standard metric through which to assess the efficacy of unsupervised clustering algorithms.}

\begin{widetext}

\begin{figure}
\begin{centering}
\includegraphics[width=0.45\linewidth]{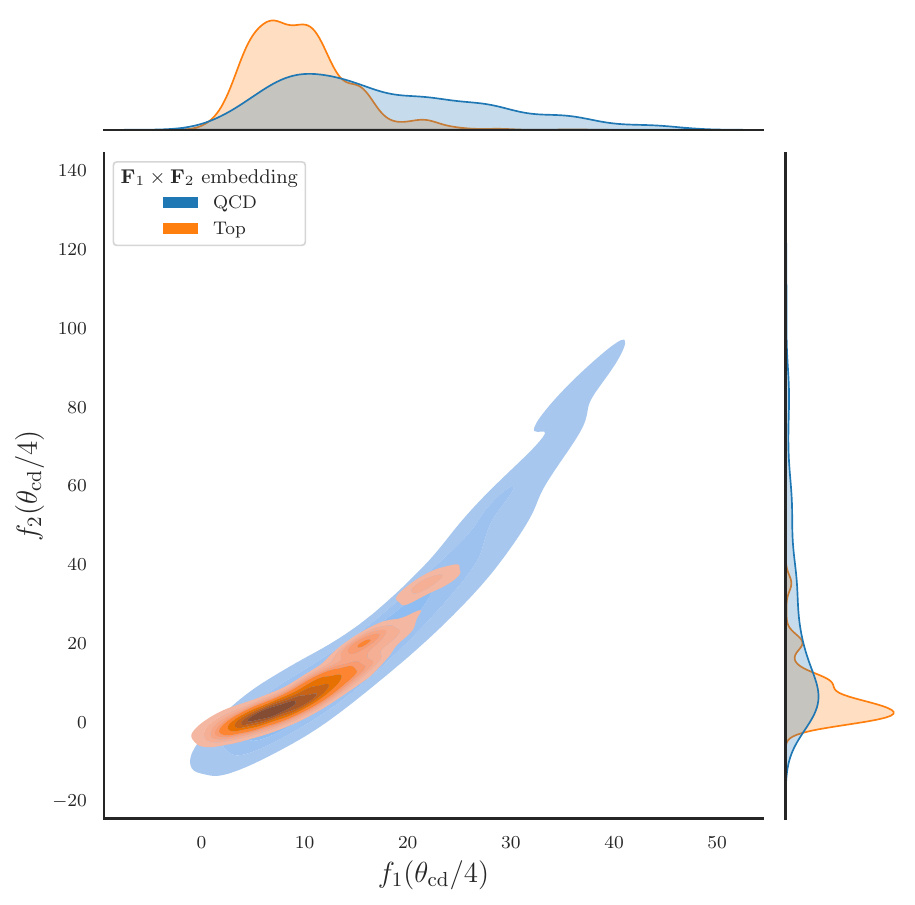}
\includegraphics[width=0.45\linewidth]{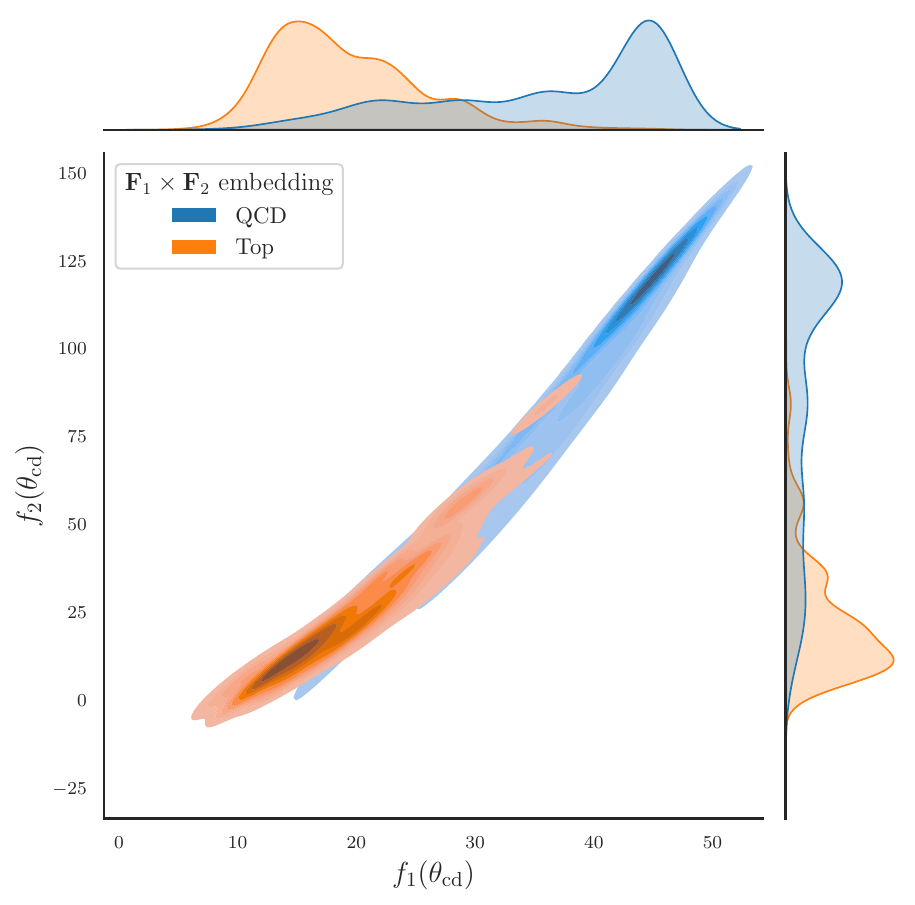}
\includegraphics[width=0.45\linewidth]{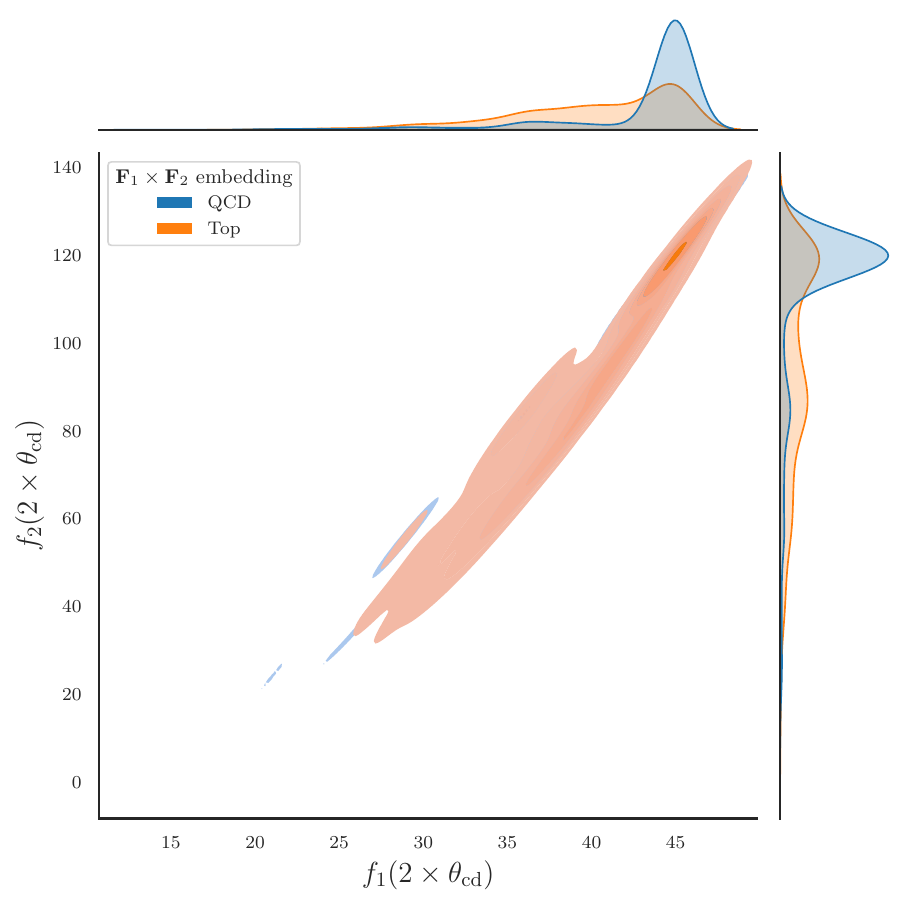}
\end{centering}
  \caption{KDE plots for the embeddings of QCD (blue) and top (orange) jets in the  $(f_1, f_2) \in \mathbf{F}_1 \times \mathbf{F}_2 \subset \mathbf{R}^2$ space at varying levels of angular resolution. The upper plots are generated at $r = \theta_{\mathrm{cd}}/4$ (left) and $r = 2\times \theta_{\mathrm{cd}}$ while the lower plot corresponds to $r = \theta_{\mathrm{cd}}$. We see the highest level of separability in the latter case.}

 \label{fig:qcd-top-kde}
\end{figure}

\begin{figure}
\begin{centering}
\includegraphics[width=0.45\linewidth]{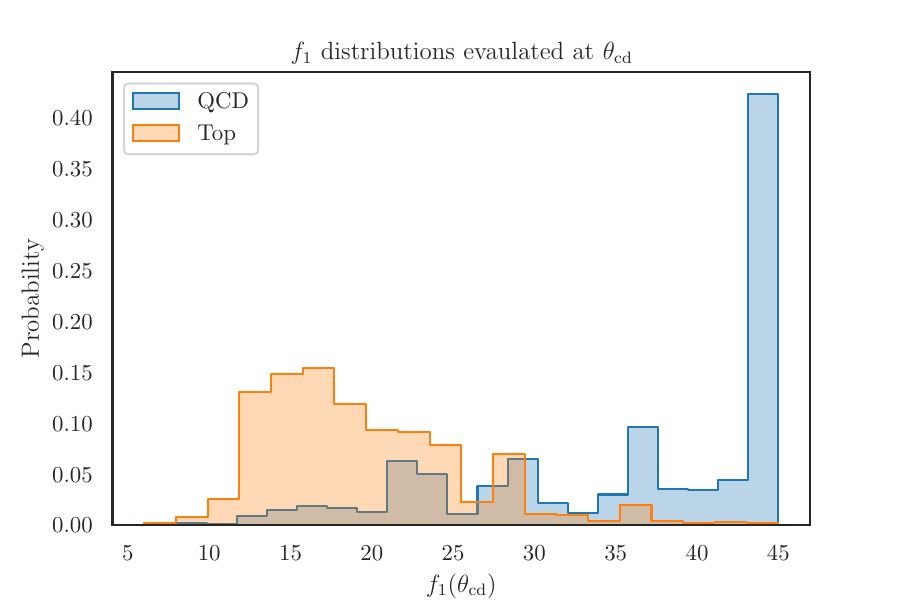}
\includegraphics[width=0.45\linewidth]{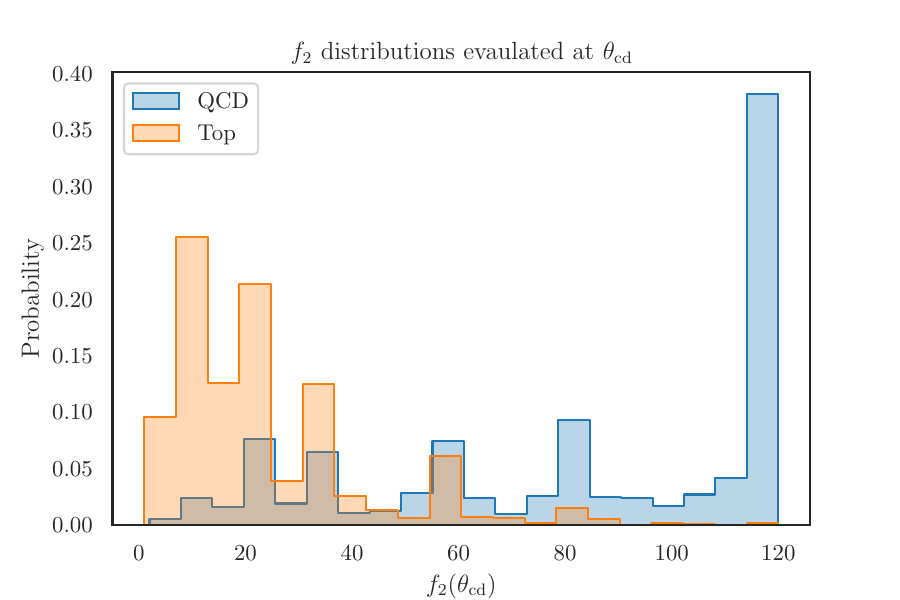}
\end{centering}
  \caption{Distributions for $f_1$ and $f_2$ at the angular resolution scale $r = \theta_{\mathrm{cd}}$. These may be viewed as projections onto the horizontal and vertical axes of the lower subfigure of Fig.~\ref{fig:qcd-top-kde}.}
 \label{fig:f1-f2}
\end{figure}

\end{widetext}

\section{Further investigations of the SFV}
\label{sec:further-investigations}

In this section, we delve further into properties of the SFV representation, but first we must develop some of the language required to do so. In what follows, we consider $f_1(r)$ and $f_2(r)$ as random variables whose stochasticity is inherited directly from that of our underlying data set. This is to say that the random fluctuations exhibited in the initial point-clouds of the QCD and top jets forming our data set are deterministically mapped to the components of the SFV through Eq.~(\ref{eq:f-def}). This allows us to analyze properties regarding probability distributions in $f_1(r)$ and $f_2(r)$, where normalization comes simply from the total number of jets in each class and class membership is denoted by the subscript $A\in \{\text{QCD, top}\}$. Thus, for each jet flavor $A$ and angular resolution $r$, there exists a one-parameter family of joint distributions in the the variables $f_1$ and $f_2$, $p_A\left(f_1,f_2 ; r\right)$.

With these considerations in mind, let us establish notation, following that of \cite{10.5555/1146355}. We will refer to components of the SFV as random variables through $F_1$ and $F_2$ that are drawn from the distribution $p_A\left(f_1,f_2 ; r\right)$--- this is denoted
\begin{align}
    F_1\,, F_2 \sim p_A\left(f_1,f_2 ; r\right)\,,
\end{align}
where
\begin{align}
    p_A\left(f_1,f_2 ; r\right) &= \mathrm{Pr}\big\{F_1 = f_1\, \mathrm{and} \,  F_2 = f_2\,,\nonumber \\
    &\hspace{1.5cm} \text{at scale} \, r\,, \text{and for flavor} \, A \big\}\,.
\end{align}
Next, we define what are known as the alphabets for $F_1$ and $F_2$, which are the discrete set of values these variables can take on. They are simply
\begin{align}
    \mathcal{F}_1 &= \left\{0,1, \dots , \binom{N_{\mathrm{jet}}}{2} \right\} \,,\nonumber \\
    \mathcal{F}_2 &= \left\{0,1, \dots , \binom{N_{\mathrm{jet}}}{3} \right\}\,.
\end{align}
In light of the previous section, we see that these discrete sets are those that are embedded into the continuum spaces, $\mathcal{F}_{1,2} \hookrightarrow \mathbb{F}_{1,2}$ and visualized in Fig.~\ref{fig:qcd-top-kde}. 

Next, the information of a given pair $(f_1, f_2)$ at scale $r$ and for jet-flavor $A$ is defined as
\begin{align}
    H_A(f_1, f_2;r) = \log \frac{1}{p_A\left(f_1,f_2 ; r\right)}\,,
\end{align}
where the intuition goes as follows. The larger the logarithm of the inverse probability to measure the pair $(f_1, f_2)$, the smaller the probability weight affiliated with such a measurement, and thus the more informative such a measurement is about the underlying distribution from whence it is sampled.\footnote{This is why this quantity is often also referred to as the ``surprise'' of a measurement.} In this work, we use the natural logarithm, or log base $e$ and therefore measure information in what are known as ``nats'' (the usual ``bits'' correspond to log base 2).

Finally, the KL divergence (or relative entropy) between two distributions $P$ and $Q$ of random variable $X$ with alphabet $\mathcal{X}$ is a measure of information gain, or inefficiency, in modeling the ``true'' distribution $P$ by the ``model'' $Q$. While the KL divergence is a generalization of a squared-distance, it is not a metric, as it is not symmetric and does not satisfy the triangle inequality \cite{10.5555/1146355}. It is computed as
\begin{align}
    D_{\rm KL} \left(P \, \lVert \, Q \right) = \sum_{x \in \mathcal{X}} P(x) \log \frac{P(x)}{Q(x)}\,.
\end{align}

These basic ingredients will be utilized in what follows. First, in Sec.~\ref{sec:variable-info} we explore some basic information-theoretic and geometric considerations regarding the random variables $F_1$ and $F_2$ themselves and considering the flavor-dependence of such considerations separately. Then in Sec.~\ref{sec:flavor-info}, we will consider the information overlap and distances between distributions as indexed by their jet flavor $A$ resulting from their underlying values of $(f_1,f_2)$. We emphasize that the $r$-dependence inherited by the SFV from the simplicial substructure complex $K_{\mathrm{sub}}(r)$ affords us one-parameter families of distributions $p_A\left(f_1,f_2 ; r\right)$, which in turn allow us to study all the following information-theoretic and geometric features as functions of the angular resolution scale $r$.

\subsection{Information and geometry of the random variables $F_1$ and $F_2$}
\label{sec:variable-info}

Upon examining Fig.~\ref{fig:qcd-top-kde}, we qualitatively see correlation between the $f_1$ and $f_2$ coordinates. Now, it is intuitively obvious that $f_1$ and $f_2$ should be correlated, and a natural information-theoretic object to quantitatively investigate such a correlation is the mutual information between the two random variables $F_1$ and $F_2$. This is defined as
\begin{align}
    &I_A \lb F_1, F_2; r\rb \nonumber \\
    &\hspace{0.5cm}= \sum_{f_1 \in \mathcal{F}_1\,,\, f_2 \in \mathcal{F}_2}p_A\lb f_1,f_2; r\rb\log \frac{p_A\lb f_1, f_2; r\rb}{p_A\lb f_1; r\rb \, p_A\lb f_2 ; r\rb}\,, 
\end{align}
or equivalently
\begin{align}
\label{eq:mutual-info}
    &I_A \lb F_1, F_2; r\rb = D_{\rm KL} \lb p_A\left( f_1, f_2 \right) \, \big\lVert \, p_A\left(f_1\right) \,p_A\left(f_2\right)\rb\,, 
\end{align}
where 
\begin{align}
    p_A\lb f_i;r \rb = \sum_{f_j \in \mathcal{F}_j}p_A \lb f_i, f_j; r \rb\,, \, \text{for } i\in \{1,2\}\,, \, j \neq i\,,
\end{align}
are the marginalized distributions. Eq.~(\ref{eq:mutual-info}) thus tells us that the mutual information quantifies the extent to which the joint distribution in $(f_1,f_2)$ deviates the product of its marginals. Such a product only well-approximates the joint distribution in the limit that $F_1$ and $F_2$ become independent.

A quantity closely-related to the mutual information is what's known as the variation of information---it is defined by
\begin{align}
\label{eq:variation-info}
    \mathrm{VI}_A\left(F_1, F_2; r\right) &= S_A\left(F_1, F_2; r \right)  - I_A\left(F_1, F_2 ; r\right)\,,
\end{align}
where $I\left(F_1, F_2 ; r\right)$ is given by Eq.~(\ref{eq:mutual-info}) and $S_A\lb F_1, F_2;r \rb$ is the joint Shannon entropy
\begin{align}
    S_A\lb F_1, F_2;r \rb = -\sum_{f_1 \in \mathcal{F}_1\,,\, f_2 \in \mathcal{F}_2} p_A\left( f_1, f_2\right)\log p_A\left(f_1, f_2 \right)\,.
\end{align}
We denote the variation of information in Eq.~(\ref{eq:variation-info}) by $\mathrm{VI}_A\left(F_1, F_2; r\right)$ because it satisfies all the requirements for a distance metric. As such, it can be interpreted as the distance between the random variables $F_1$ and $F_2$, or in other words, the separation in the information contained in $F_1$ and $F_2$, at the scale $r$

\begin{figure}
\begin{centering}
\includegraphics[width=\linewidth]{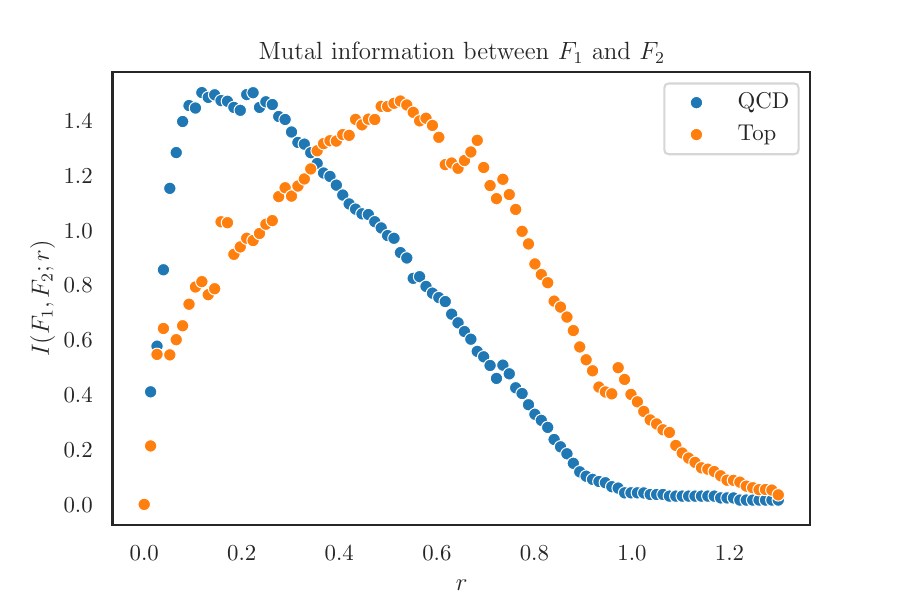}
\end{centering}
\caption{Mutual information for QCD (blue) and top (orange) jets. The mutual information captures in the information overlap between the random variables $F_1$ and $F_2$.}
\label{fig:qcd-top-mutual-info}
\end{figure}

\begin{figure}
\begin{centering}
\includegraphics[width=\linewidth]{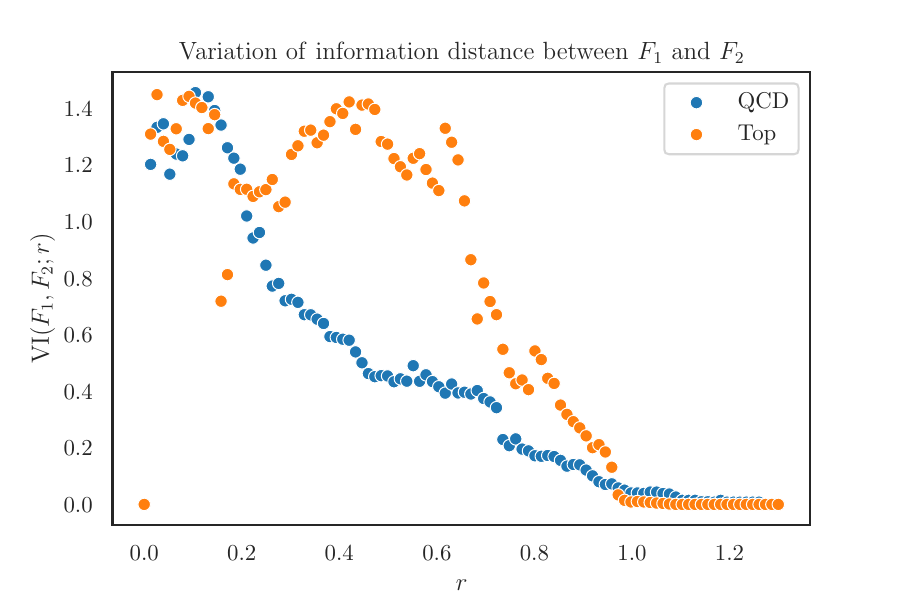}
\end{centering}
 \caption{Variation of information for QCD (blue) and top (orange) jets. The variation of information is a measure of the distance between the random variables $F_1$ and $F_2$.}
 \label{fig:qcd-top-variation of info}
\end{figure}

With these objects defined, we may see how they behave as functions of the angular resolution for QCD and top jets separately, beginning with the mutual information displayed in Fig.~\ref{fig:qcd-top-mutual-info}. Based on the discussion surrounding Eq.~(\ref{eq:mutual-info}), we may identify the regions in which $I_A\left(F_1, F_2 ; r\right)$ reaches a global/local maximum as the regions in which the $F_1$ and $F_2$ variables have the highest correlation. These are the very regions in which both 1- and 2-simplexes proliferate concurrently, and such regions are those where dense clusters of particles are to be found, as each $B_i(r)$ has nonzero intersection with neighboring balls. Hence these are the regions in which the angular scale $r$ resolves dense clusters. 

Now, looking to Fig.~\ref{fig:qcd-top-mutual-info}, we may understand how this behavior manifests itself for both QCD and top jets. First, in the QCD case, we identify a maximum plateau for $r \sim [0.1, 0.2]$ after which $I_{\rm QCD}\left(F_1, F_2 ; r\right)$ falls steadily---thus identifying a dense core of collinear radiation about a single hard prong. This type of behavior is only experienced in $I_{\rm top}\left(F_1, F_2 ; r\right)$ once $r \gtrsim \theta_{\mathrm{cd}}$. This is a byproduct of the extended structure of top-initiated jets, due to the top quark's characteristic hard decay pattern of $t \rightarrow q \bar{q}^\prime b$, whose extent is only fully-resolved for these high values of $r$. Thus, we see that in both cases the mutual information provides a gauge for the effective angular width of a jet's substructure, for it tracks the level of correlation between the proliferation of 1- and 2-simplexes. Such proliferation proceeds once the bulk of the particles are resolved, and then slowly decays down to zero once the values of $f_1$ and $f_2$ approach their limits. Alternatively, we may interpret the $r$-dependent mutual information to provide a proxy for the density with which particles fill their respective jets. The lower the value of $r$ for which $I_A\left(F_1, F_2 ; r\right)$ attains its max, the more highly-collimated and dense the jets are, and vice versa.

The variation of information, $\mathrm{VI}_A\left(F_1, F_2; r\right)$, is shown in Fig.~\ref{fig:qcd-top-variation of info}. Due to its close relation to $I_A \lb F_1, F_2; r\rb$, we may reasonably expect qualitatively similar behavior to exhibit itself, however, we see that this metric reveals some interesting sharper features. First, let us note the features this metric has in common with the mutual information, which are the approximate locations of the maximum for QCD jets near $r\sim 0.1$ and the max plateau region for top at $r\gtrsim \theta_{\mathrm{cd}}$. 

The feature that is most striking is the emergence of two maxima of the same size for the case of top jets. Considering Figs.~\ref{fig:qcd-top-mutual-info} and ~\ref{fig:qcd-top-variation of info} together and inspecting Eq.~(\ref{eq:variation-info}), we see that this additional maximum originates in a local maximum in the joint entropy $S_{\mathrm{top}}\lb F_1,F_2;r \rb$, which must attain a lower value than its global max---we see that global maximum must occur at the same location as $I_{\mathrm{top}}\lb F_1,F_2;r \rb$ with value about twice that of the mutual information. The presence of this local max is due to the proliferation of simplexes in the smaller sub-clusters of particles that appear in top jets. We know these must form in subsets of the total number of particles contained in the jet, because the totality of the top jet's constituents can only be resolved at the higher values of $r$.

Now, what makes the variation of information a particularly interesting metric is the physical interpretation of the peaks near $r \sim 0.1$ for both QCD and top jets. We can interpret these peaks to define the resolution scale for collinear radiation about hard colored prongs due to parton showering---the single prong for QCD jets and the triplet in the case of tops. Remarkably, we have
\begin{align}
    S_{\mathrm{QCD}}\lb F_1,F_2;0.1 \rb \approx S_{\mathrm{top}}\lb F_1,F_2;0.1 \rb\,,
\end{align}
despite the two jet classes resolving widely differing numbers of particles in their respective clusters---$\mathcal{O}(10)$ for QCD and $\mathcal{O}(1)$ for top. This implies that the distance between $F_1$ and $F_2$ peaks at angular scales resolving subjets. This is then further corroborated by 
\begin{align}
    S_{\mathrm{top}}\lb F_1,F_2;0.1 \rb \approx S_{\mathrm{top}}\lb F_1,F_2; 0.5 \rb\,,
\end{align}
where for $r\sim 0.5$, the balls surrounding particles contained in each of the three prongs of the top jets have large enough radii to form simplexes with those of neighboring prongs, and we achieve enter the regime of maximal joint entropy. This regime can then be interpreted as the regime where $r$ is too large to resolve the details of the particles forming each prong, but can now recognize the emergence of a new set of subjets to recluster and subsume into a larger one. Thus, we see the maxima of $\mathrm{VI}_A\left(F_1, F_2; r\right)$ as indicators of the telescoping/fractal structure of jets. While QCD jets are scale-free and thus contain only one peak denoting continual self-similarity as $r$ grows, the presence of the top mass introduces an additional scale, thereby introducing an additional peak in $\mathrm{VI}_A\left(F_1, F_2; r\right)$, and hence signifies an additional layer of substructure through which to cluster subjets contained within top jets.

The fundamental differences between QCD and top jets are thus revealed quite dramatically by the SFV data type. In the following section, we demonstrate that the sensitivity of the SFV to absence/presence of extended structures may be leveraged to reveal distinguishing distinguishing features of QCD and top jets.

\subsection{Information and geometry of QCD and top jet distributions}
\label{sec:flavor-info}

\begin{figure}
\begin{centering}
\includegraphics[width=\linewidth]{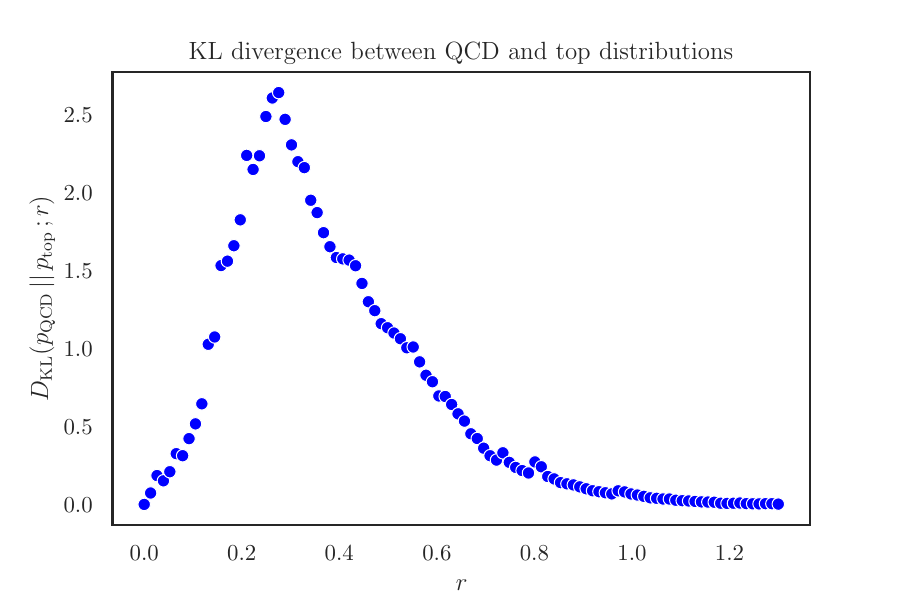}
\end{centering}
  \caption{The KL divergence between QCD and top $f_1$-$f_2$-distributions. The KL divergence captures in the information-theoretic surprise one gets in modeling the top distribution by that of the QCD.}
 \label{fig:KL}
\end{figure}

In this section, we investigate how the SFV data type can be used to explicitly differentiate QCD and top jets. Such differentiation is the focus of many ML-based studies in the HEP literature, as top-tagging is a task of primary interest at colliders such as the LHC. The study of flavor relations between distributions can be cast into the physical language of top-tagging, as such a pursuit consists of one searching a background high-probability haze of QCD jets for anomalous low-probability top jets. Recalling our previous discussion, the KL divergence
\begin{align}
    &D_{\rm KL} \left(p_{\rm QCD} \, \lVert \, p_{\rm top}; r\right) \nonumber \\
    &\hspace{0.25cm}= \sum_{f_1 \in \mathcal{F}_1\,,\, f_2 \in \mathcal{F}_2} p_{\rm QCD}\left(f_1, f_2;r \right) \log \frac{p_{\rm QCD}\left(f_1, f_2  ;r\right)}{p_{\rm top}\left(f_1, f_2 ;r\right)} \,,
\end{align}
measures the inefficiency of modeling the distribution of background QCD jets by the distribution of the rare top jets. The resulting curve is displayed in Fig.~(\ref{fig:KL}), where we see that the KL divergence is maximized sharply for $r= \theta_{\mathrm{cd}}$. Thus, this quantity readily identifies the angular scale unique to tops.

What is perhaps more interesting is the location of the peak in the KL divergence. We note that our data set contains jets of $p_T^{\rm jet} \in [550, 650]$ GeV, and thus we see that 
\begin{align}
    \argmax_r  D_{\rm KL} \left(p_{\rm QCD} \, \lVert \, p_{\rm top}; r \right) \approx  \theta_{\rm cd} \,.
\end{align}

\begin{figure}
\begin{centering}
\includegraphics[width=\linewidth]{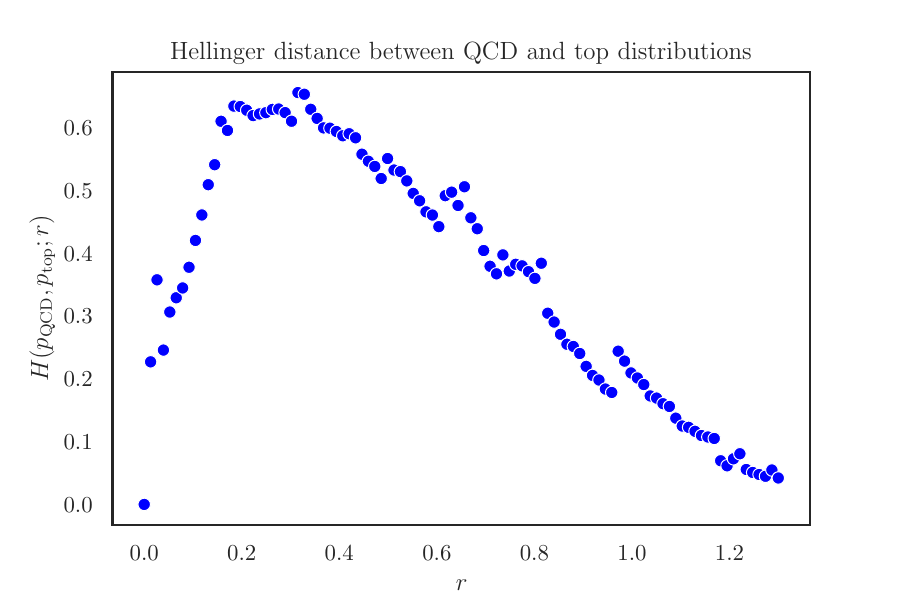}
\end{centering}
 \caption{The Hellinger distance between QCD and top $f_1$-$f_2$-distributions. The Hellinger distance defines a metric on between these distributions.}
 \label{fig:qcd-top-hellinger}
\end{figure}

While the KL divergence is a useful measure of differing information content for the two jet flavors, such analysis may be complemented by the study of a metric defined on the distributions themselves. One such metric is the Hellinger distance. For the case of QCD and top SFV distributions, the square of this distance is computed as
\begin{align}
    &H^2(p_{\mathrm{QCD}}, p_{\mathrm{top}}; r)  \nonumber \\
    &= \frac{1}{2}\sum_{f_1 \in \mathcal{F}_1, f_2 \in \mathcal{F}_2}  
 \biggl( \sqrt{p_{\mathrm{QCD}}(f_1,f_2; r)} - \sqrt{p_{\mathrm{top}}(f_1,f_2; r)} \biggr)^2.
\end{align}
$H(p_{\mathrm{QCD}}, p_{\mathrm{top}}; r)$ is displayed in Fig.~\ref{fig:qcd-top-hellinger}. The unique feature of the Hellinger distance is its effective identification of the two aforementioned angular resolution scales that appear in top jets. Note that Hellinger distance displays a single global maximum plateau defined sharply by the region $r\in [0.1, \theta_{\rm cd}]$. This highlights something that is seen neither in the analysis of Sec.~\ref{sec:variable-info}, nor in that of the KL divergence of the present section. The fact that the angular scale $r \approx 0.1$ marks a boundary for the maximal distance between distributions means that at the level of collinear radiation, the QCD and top distributions differ sharply in their values of $f_1(0.1)$ and $f_2(0.1)$, as at this scale, QCD jets have nearly the entirety of their particles resolved, whereas top jets have only the particles immediately surrounding each core resolved. Thus, QCD jets exist in their maximal delta function regime, whereas top jets still require a wider angular scale to capture the entirety of their particle content. This latter point is then the reason that the upper boundary of the Hellinger maximum lies at $\theta_{\rm cd}$---for once this scale is reached, neighboring cores are able to resolve each other's presence, and the top distribution thus begins its limiting to the delta function regime. Thus, the Hellinger distance is the best means of understanding the separability of QCD and top jets in $\mathbb{R}^2_{\rm S}$ shown in 
Fig.~\ref{fig:qcd-top-kde}.

\section{Simplicial jet shape observable}
\label{sec:jet-shape}

In this final section, we define an experimental observable that follows naturally from the SFV data type. This observable is closely related to what are known as the integrated and differential jet shapes and is defined nearly identically.

The integrated jet shape is a classic jet substructure observable and can be understood as follows. An axis within the jet is chosen about which an angular scale $r$ is chosen. The energies of the particles contained within the solid disk spanned by $r$ are summed up and normalized by the overall energy of the jet. It may thus be interpreted as a cumulative distribution function in the angle $r$ obtained through the first moment of the particle-energy-fraction, $z$, distribution. Thus letting $z$ denote the fraction of energy carried within the radius $r$, so that $z$ has implicit dependence on $r$, we have
\begin{align}
\label{eq:classic-shape-int}
    \Psi(r) = \int_0^1 dz \, z(r) \, p(z(r))\,,
\end{align}
where $p(z(r))$ is the distribution of energy fractions $z$ evaluated at the resolution $r$ of a sample of jets. The differential jet shape is then simply the $r$-derivative of Eq.~(\ref{eq:classic-shape-int}), giving it the interpretation of a probability density function with respect to $r$:
\begin{align}
\label{eq:classic-shape-diff}
    \psi (r) = \frac{d}{dr} \Psi(r)\,,
\end{align}
see \cite{Seymour_1997, Seymour_1998, Vitev_2008, Chien_2014, Chien_2016, Kang:2017mda, Neill:2018wtk, Cal_2019} for various theoretical calculations of this observable.

We see that this definition lends itself quite naturally to an analogous jet shape observable with respect to the SFV. Here $\bs f(r)$ plays the role of $z(r)$ which is the accumulated number of simplexes contained within the angular resolution variable $r$. A key difference, though, is that $\bs f(r)$ makes no reference to a particular axis choice, as the angular scale $r$ is that which defines balls about each particle contained within the jet. Thus, the pre-processing independence of the SFV leads to an independence in the choice of axis for the simplicial shape.\footnote{From a theoretical standpoint, computation of the traditional jet shape has a remarkably strong dependence on the choice of axis through remarkably different QCD factorization theorems that each axis requires \cite{Kang:2017mda}.} Let $\lVert \bs f(r) \rVert = \sqrt{f_1(r)^2 + f_2(r)^2} \equiv \mathcal{N}(f_1, f_2; r) $, and then, in analogy with Eq.~(\ref{eq:classic-shape-int}), we define the integrated simplicial shape to be
\begin{align}
\label{eq:simp-shape-int}
    \Sigma (r) = \sum_{f_1\in \mathcal{F}_1\,,\, f_2 \in \mathcal{F}_2} \mathcal{N}(f_1,f_2; r) \, p \lb f_1, f_2; r \rb\,,
\end{align}
so that the differential version is directly analogous to Eq.~(\ref{eq:classic-shape-diff}), i.e.
\begin{align}
    \sigma(r) = \frac{d}{dr} \Sigma (r)\,.
\end{align}
Note that the cumulative distribution (probability density) function interpretations of the integrated (differential) shape is not being used here, as we are taking $\mathcal{N}(f_1,f_2; r)$ to not be normalized by its maximum value. Our choice is simply one of many choices for simplicial shape observables. 

\begin{figure}
\begin{centering}
\includegraphics[width=\linewidth]{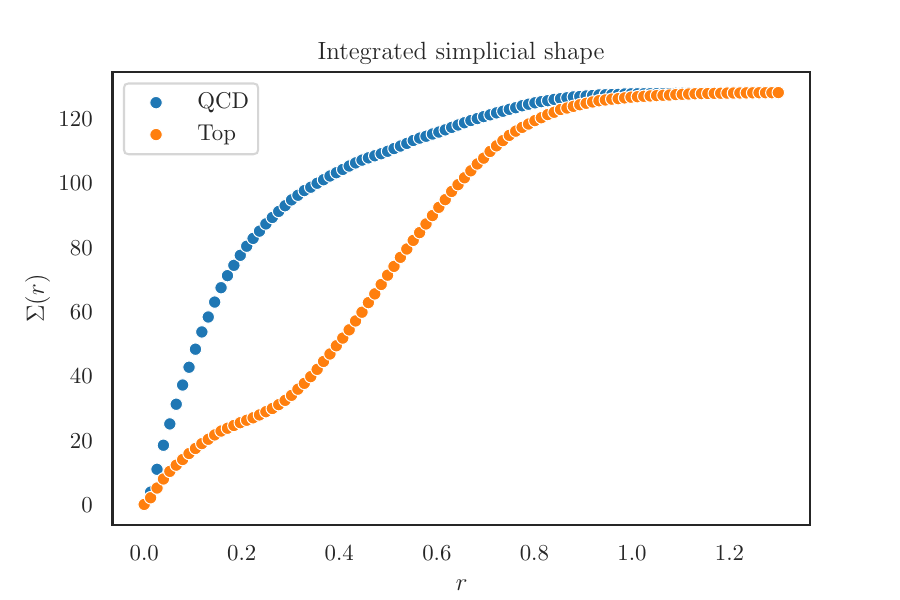}
\end{centering}
 \caption{Integrated simplicial shapes for QCD (blue) and top (orange) jets.}
 \label{fig:int-shape}
\end{figure}

\begin{figure}
\begin{centering}
\includegraphics[width=\linewidth]{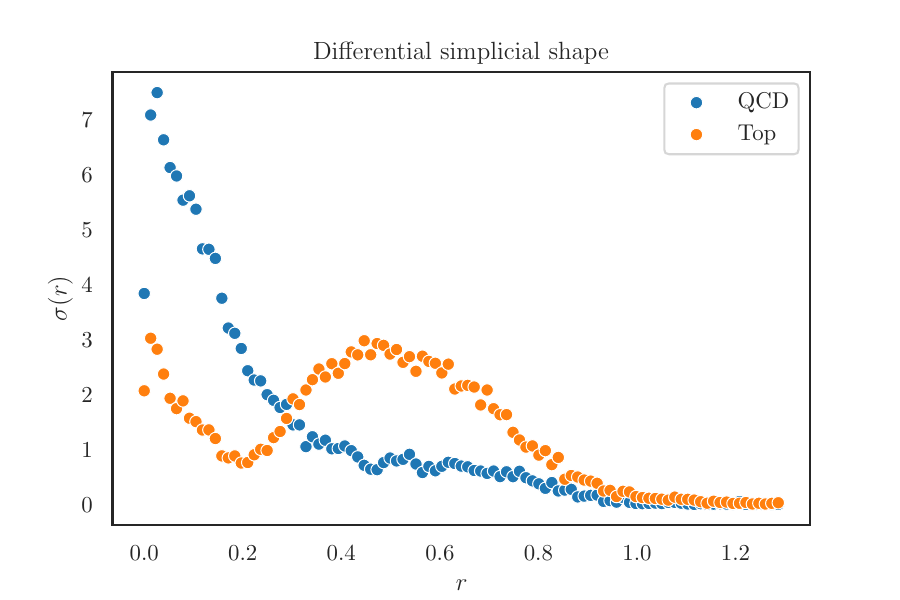}
\end{centering}
 \caption{Differential simplicial shapes for QCD (blue) and top (orange) jets.}
 \label{fig:diff-shape}
\end{figure}

Plots for $\Sigma(r)$ and $\sigma(r)$ for both QCD and top jets are displayed in Figs.~\ref{fig:int-shape} and \ref{fig:diff-shape}, respectively. They display behavior that is congruent with the discussions of Sec.~\ref{sec:further-investigations}. At both the integrated and differential levels, we see the rapid accumulation of simplexes at the resolution scale $r\sim 0.1$ for QCD jets, while for top jets, this takes place initially at $r\sim 0.1$ where the QCD radiation surrounding each prong is resolved, and then again once $r\gtrsim \theta_{\rm cd}$ and where the prongs can be resolved into a single large-scale cluster. Thus both $\Sigma(r)$ and $\sigma(r)$ are shape observables which are particularly well-suited for exhibiting the extended nature of decay patterns within the substructure of a jet. We remark here that the shapes displayed are simply for the particular choice of $\mathcal{N}(f_1, f_2;r)$ given above, and that there certainly exist many more functions of SFV components one could choose---and in fact it would be very interesting so see if different choices are better suited for identifying other important physical features of a jet's substructure. On such a note, it would also furthermore be interesting to see the corresponding profiles for jets whose progeny comes from other heavy resonances, such as $W^{\pm}$, $Z$, or $H$ bosons, for the identification of their respective extended structures could serve in the performance of precision measurements of the masses of such particles.

\section{conclusion}
\label{sec:conclusion}

In this work, we have proposed a fundamentally new data type for the representation of jets, which we refer to as the simplicial substructure complex, $K_{\mathrm{sub}}(r)$. This data type is the union of sets of 0-, 1-, and 2-simplexes, whose elements are dependent on the internal angular resolution variable $r$ of a given jet. This data type gives rise to two natural representations, namely the graph representation $G_{\mathrm{sub}}(r)$ and the SFV representation $\boldsymbol{f}(r)$. Each representation allows for the computation of novel features. The graph representation allows for the insights of graph theory to be applied to jet substructure, which is particularly useful in the study of the prong structure of jets. The SFV representation readily lends itself to visualization, analytic computation of distances between vectors, as well as a host of information-theoretic and geometric considerations at the level of random variables and distribution flavors. 

Thus, we see that the SFV distinguishes itself in many previously-mentioned regards, but we highlight one in particular---that is its scale-dependence. The analysis of such dependence is of central interest to both experimental and theoretical investigations of jet substructure and is a physically-natural dial through which to probe the internal workings of jets. This scale-dependence is also of fundamental importance in the field of topological data analysis, which certainly makes TDA seem to be a natural lens through which to view jet substructure. We hope that this work can serve to open the door to a greater union between these disciplines as well as many more interesting studies along this line.

\section{Acknowledgments}
The authors thank A.J. Larkoski for helpful and interesting discussions. The work of JR is supported by the Mani L. Bhaumik Institute for Theoretical Physics.

\bibliographystyle{bibstyle}
\bibliography{metric}

\end{document}